\documentclass[journal=jacs,manuscript=article]{achemso}

\SectionNumbersOn

\usepackage[version=3]{mhchem} 
\usepackage[T1]{fontenc} 
\usepackage{booktabs}
\usepackage{color}
\usepackage{hyperref}
\usepackage{enumerate}
\usepackage{amsmath}
\usepackage{cleveref} 
\usepackage[final]{pdfpages}

\author{Erik  D. Hedeg{\aa}rd}
\affiliation[Lunds Universitet]
{Department of Theoretical Chemistry, Lund University, P. O. Box 124, SE-221 00 Lund, Sweden}
\email{erik.hedegard@teokem.lu.se}
\author{Ulf Ryde}
\affiliation[Lunds Universitet]
{Department of Theoretical Chemistry, Lund University, P. O. Box 124, SE-221 00 Lund, Sweden}
\email{ulf.ryde@teokem.lu.se}

\title{Multiscale Modelling of Lytic Polysaccharide Monooxygenases}

\begin{document}







\begin{abstract}

Lytic polysaccharide monooxygenase (LPMO) enzymes have attracted considerable attention, owing to their ability to 
enhance polysaccharide depolymerization, making them interesting in respect to production of biofuel from cellulose. The LPMOs 
are metalloenzymes that contain a mononuclear copper active site, capable of activating dioxygen. However, many details of this activation are unclear. Some 
aspects of the mechanism have previously been investigated from a computational angle. Yet, these studies have either employed only  
molecular mechanics (MM), which are inaccurate for metal active sites, or they have described only the active site with quantum mechanics (QM) 
and neglected the effect of the protein. Here, we employ hybrid QM and MM (QM/MM) methods 
to investigate the first steps of the LPMO mechanism, which is reduction of of Cu$^{\text{II}}$  to Cu$^{\text{I}}$ 
and formation of a Cu$^{\text{II}}$--superoxide complex. In the latter complex, the superoxide can bind either in an equatorial or an axial position. 
For both steps we obtain structures that 
are markedly different from previous suggestions, based on small QM-cluster calculations. Our calculations show that the equatorial isomer of the superoxide complex is  
over 60 kJ/mol more stable than the axial isomer, being stabilized by interactions with a second-coordination-sphere Gln residue, showing a possible role for this residue. 
Coordination of superoxide in this manner is in agreement with recent experimental suggestions.   

\end{abstract}

\section{Introduction}\label{Introduction}

Employing cellulose in biofuel production can make this advancing technology a highly competitive alternative to fossil fuel. 
As a major component of biomass, cellulose is non-expensive and the most abundant polysaccharide on earth\cite{klemm2005}. 
However,  the application of cellulose in biofuel production requires degradation of it into smaller sugars, which 
has shown to be a major obstacle, requiring both hydrolytic enzymes and thermal work. 
This remarkable  stability is caused by the structure, involving very long glucan chains, which are comprised of 
glucose monomers, coupled together by $\beta$-1,4 glycosidic linkages. The glucan chains interact with each other by an extensive network of inter- and 
intra-molecular hydrogen bonds invovling the three hydroxyl groups of each glucose monomer,  
which limits the accessibility of hydrolytic enzymes to the glycosidic linkages.\cite{chang2007,himmel2007}  

 Lately, a class of  fungal and bacterial enzymes called 
lytic polysaccharide monooxygenases (LPMOs) have attracted considerable attention owing to their ability to 
enhance polysaccharide depolymerization, thereby providing a route to efficient 
conversion of cellulose into fermentable sugars\cite{hemsworth2013a,beeson2015,span2015}. The first industrial applications have 
already been seen for ethanol production.\cite{harris2014} 
\begin{figure}  
\includegraphics[width=0.75\textwidth]{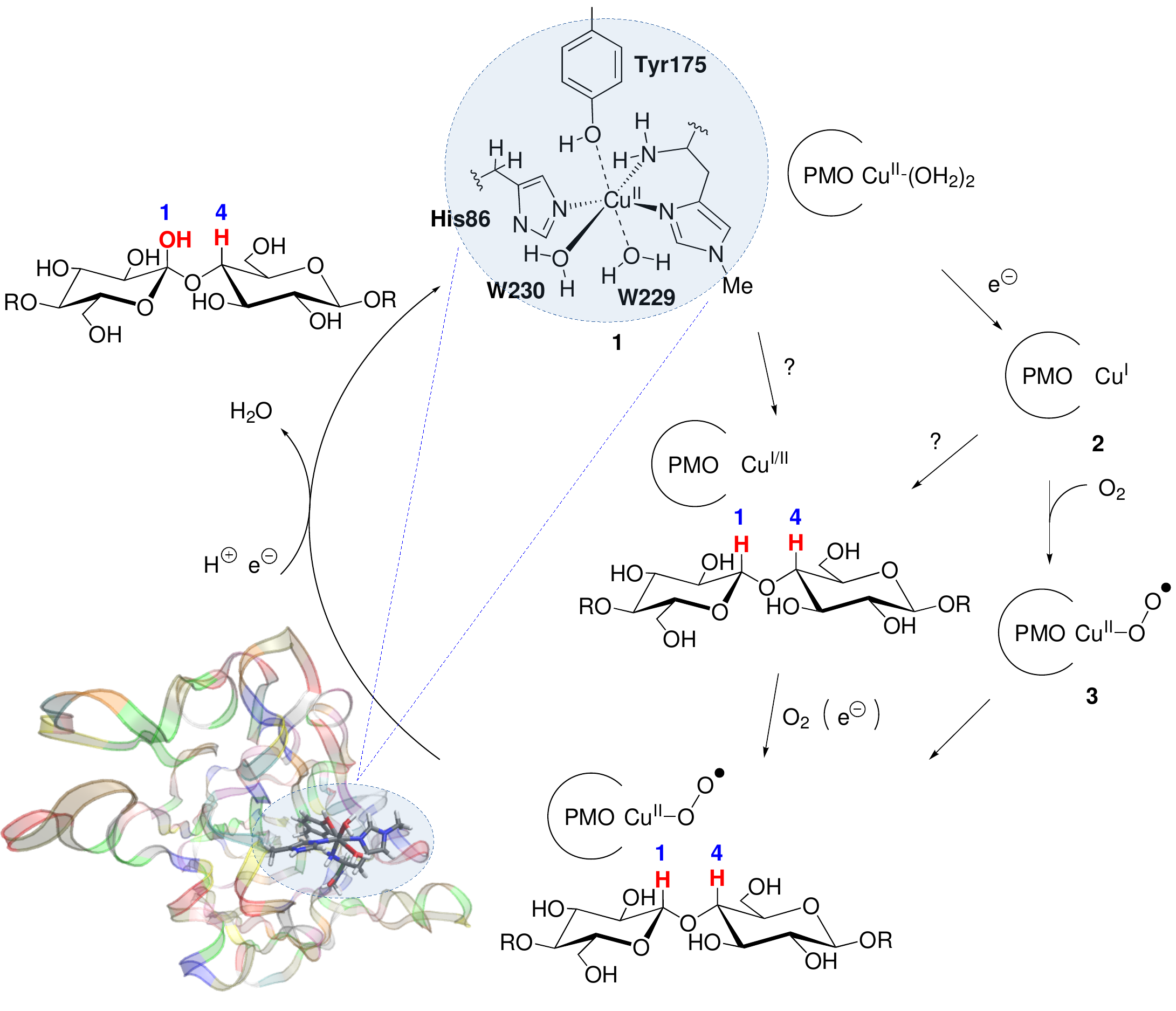}  
\caption{Active site and putative mechanism of a fungal LPMO (bacterial LPMOs lack the axial tyrosine residue, but are likely to work through a similar mechanism). 
The order of events (electron transfer, substrate uptake and \ce{O2} coordination) is not clarified 
as suggested by the different possible paths. Residue numbers refer to the enzyme from \textit{Thermoascus aurantiacus} (PBD 2YET).\cite{quinlan2011}  \label{lpmo_active_site}}   
\end{figure} 
Initially, LPMO enzymes\cite{raguz1992,armesilla1994} were assumed to be solely hydrolytic and were classified 
 as belonging to the glycoside hydrolase 61 (GH61) and carbohydrate-binding modules 33 (CBM33) families. Initial reports questioning whether  GH61 indeed were hydrolases occurred 
in 2008\cite{karkehabadi2008} and in 2010 Harris et al.\cite{harris2010} showed that a LPMO from the GH61 family significantly enhanced cellulase activity 
(a few earlier reports are also known from the patent literature\cite{dotson2007,brown2008}).  
The exact role of the LPMOs was demonstrated later in 2010 by Vaarje-Kolstad et al.\cite{vaaje-kolstad2010} who 
showed that a bacterial LPMO belonging to the CBM33 family exhibited an oxidative mechanism.
Importantly, these studies suggested involvement of a metal, 
although the nature of this metal was not immediately elucidated.  Yet, it was clear that the LPMOs employed a common mechanism, different  
from traditional glycoside hydrolases, and the enzymes are today reclassified as auxiliary activity enzymes 
AA9 (formerly GH61) and AA10 (formerly CBM33).\cite{levasseur2013}. Quinlan et al.\cite{quinlan2011} could from X-ray crystallography and electron paramagnetic resonance (EPR) spectroscopy
show that LPMOs employ copper in the active site, which solved the initial 
confusion concerning the metal site\cite{karkehabadi2008,harris2010,vaaje-kolstad2010}. Later, several other reports 
confirmed that both AA9 and AA10 enzymes contain a 
single copper ion\cite{quinlan2011,phillips2011,beeson2012,horn2012,forsberg2011,hemsworth2013b,vaaje-kolstad2012,aachmann2012,vaaje-kolstad2013}.  

The active site of an AA9 enzyme is shown in \ref{lpmo_active_site}, together with a putative mechanism.\cite{phillips2011} 
For the resting state, \textbf{1}, the first coordination sphere is distorted octahedral and comprised of a weakly coordinated tyrosine, 
two water molecules and an unusual \textit{histidine brace}
comprised of two histidine residues, coordinating through the imidazole group. One of these is the N-terminal residue, which also 
coordinates through the terminal \ce{NH2} group\cite{quinlan2011}, as also seen in the particulate methane monooxygenases.\cite{solomon2014} 
Notably, a peculiar methylation of this N-terminal histidine has been observed 
in many fungal LPMO structures\cite{karkehabadi2008,westereng2011,quinlan2011,li2012,bey2013,wu2013}, although its role is unclear, as 
the non-methylated enzyme is also catalytically active\cite{wu2013,bey2013}.

EPR spectroscopy\cite{quinlan2011} have shown that \textbf{1} contains a Cu$^{\text{II}}$ ion\cite{sommerhalter2005}. 
The EPR spectra provide an important complement to X-ray structures, because Cu is known to be photo-reduced in the X-ray beam. 
In fact, it is likely that most LPMO crystal structures have contained a mixture of Cu$^{\text{II}}$ and Cu$^{\text{I}}$\cite{wu2013,hemsworth2013a,beeson2015}. 
The reduction gives rise to a lowering of the coordination number and many AA10 structures have been reported with an approximate T-shaped 
coordination environment indicating that the metal site contains a Cu$^{\text{I}}$ ion.\cite{hemsworth2013a,hemsworth2013b,gudmundsson2014}  

The mechanism of the LPMOs is far from clarified and even the order of reduction, substrate binding and \ce{O2} uptake is unknown, 
as is suggested by \ref{lpmo_active_site}. A complicating factor is that 
there is much sequence variation within the LPMO family\cite{tian2009,berka2011} and different LPMOs might employ slightly different mechanisms. 
For instance, 
the axial tyrosine (Tyr175 in \ref{lpmo_active_site}) is replaced by phenylalanine in most AA10 LPMOs.  
 Different mechanism could explain the differing substrate preferences among the LPMOs: 
Fungal AA9 enzymes oxidize cellulose\cite{harris2010,quinlan2011,phillips2011,beeson2012}, 
whereas AA10 enzymes oxidize both cellulose\cite{forsberg2011,forsberg2014a,forsberg2014b} and chitin\cite{vaaje-kolstad2010,forsberg2011,forsberg2014a}. 
Fungal LPMOs that break down chitin or starch have also been identified and are classified as AA11\cite{hemsworth2014} and AA13\cite{vu2014a,leggio2015}, respectively.    

Apart from varying
preferences with respect to substrate, LPMOs have also different regioselectivity: Some oxidize only the C1 atom of the glycoside linkage, others oxidze only the C4 atom, whereas still others 
can oxidize both C1 and C4. This has led to a sub-classification of the AA9 family, in that enzymes belonging to the LPMO type 1 subfamily are C1 specific, those belonging to LPMO type 2 
are C4 specific, and those belonging to LPMO type 3 oxidize both C1 and C4\cite{vu2014b}.  
Bacterial LPMOs (AA10) were initially thought to be C1 specific\cite{vaaje-kolstad2010,forsberg2011,forsberg2014b} until a C4 specific counterexample 
was identified.\cite{forsberg2014a} 

While the insight gained from experimental studies has been paramount, it has often been complemented by computational chemistry. 
For instance, the interaction between a fungal LPMO and cellulose has been studied 
with molecular dynamics (MD) methods\cite{wu2013,borisova2015}, whereas the structural change 
upon Cu$^{\text{II}}$ reduction (\textbf{1}$\rightarrow$\textbf{2} in \ref{lpmo_active_site}) 
has been studied with X-ray diffraction and X-ray absorption, combined with density functional theory (DFT)\cite{gudmundsson2014,kjaergaard2014}. 
The copper superoxide intermediate (\textbf{3}) has also been the target of multiple combined theoretical and experimental studies and 
 has played a central role in mechanistic suggestions so far. 
This intermediate is believed to either directly abstract a 
hydrogen from the substrate\cite{beeson2015} or 
function as precursor for a more reactive copper--oxyl species that is involved in hydrogen abstraction\cite{kim2014}.  

The coordination of \ce{O2} to Cu in \textbf{3} may give rise to two  isomers, 
because there are two possible coordination sites for \ce{O2} 
(although this complication has often been overlooked). One study\cite{kim2014} suggested 
that \ce{O2} binds in an axial position, trans to Tyr175, replacing W229 (cf.~\ref{lpmo_active_site}), whereas a combined spectroscopic and computational study 
suggested that \ce{O2} instead binds equatorially\cite{kjaergaard2014}, replacing W230. However, each study investigated only one of the two isomers. Further, 
all previous computation studies on LPMOs have been carried out with small models of the active site\cite{kim2014,gudmundsson2014,kjaergaard2014}, although it has been shown that 
 the protein environment is crucial for accurate structures and energetics 
in studies on transition-metal enzymes\cite{hu2009,hu2011,hu2013,sumner2013,karasulu2013,quesne2014,finkelmann2014,hedegard2015c,cortopassi2015}. 

 In this study, we investigate the reduction of the LPMO active site from \textbf{1} to \textbf{2}, as well as the two isomers of \textbf{3}. We include 
the protein environment with  the combined quantum mechanics and molecular mechanics (QM/MM) methodology.\cite{warshel1976,senn2009,visser2014,ryde2016} With our QM/MM protocol 
we can show that the small cluster models of the active site employed previously can give rise to large errors in the obtained structures.   

\section{Results}  \label{results}

We start this section by discussing 
wheather our computational protocol can reproduce the observed\cite{kjaergaard2014,gudmundsson2014} decrease in coordination number when 
Cu$^{\text{II}}$ is reduced to Cu$^{\text{I}}$ (Section \ref{rest_state}). We then proceed to discuss the \ce{O2} bound-states (Section \ref{CuO2-states}).  

\subsection{Structural changes when the resting state is reduced} \label{rest_state}

\begin{figure}[!htb]
        \centering
        \includegraphics[scale=0.70]{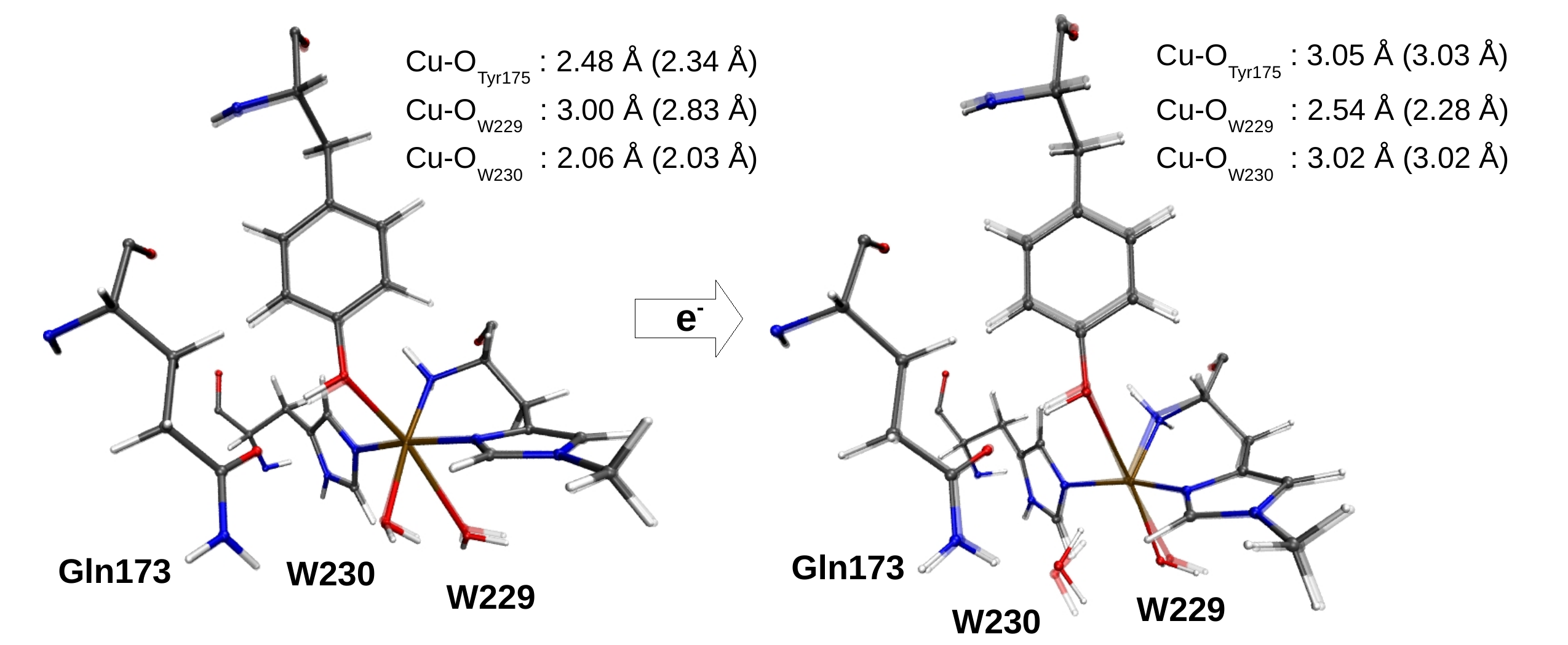}
 \caption{Structural changes upon reduction of \textbf{1} to \textbf{2}. Both figures contain an overlay of structures obtained with the def2-SV(P) (transparent) and 
def2-TZVPD basis sets. The selected bond-distances are for the def2-TZVPD basis set, with those of the def2-SV(P) basis set in parantheses. 
The second-sphere Gln173 residue is also included. \label{rest_state_vs_rest_state_red_2} } 
\end{figure}

The optimised structures and selected Cu--ligand distances of \textbf{1} and \textbf{2} are shown in \ref{rest_state_vs_rest_state_red_2}. The bond distances are compared with 
 computational results from the literature and with bond distances from a number of crystal structures in \ref{opt_rest_state}.  
\begin{table*}[htb!]
\centering
\caption{Cu--ligand bond lengths (\AA) for the active site of LPMO. \label{opt_rest_state} }
\begin{tabular}{llccccccc}
 \hline\hline \\[-2.0ex]
\textbf{State}               & \ce{Cu^{n}}  (Spin)            &  \ce{Cu-N$^{\epsilon}$_{His86}} &  \ce{Cu-N}$_{\text{His1}}$  & \ce{Cu-N}$^{\delta}_{\text{His1}}$  &  \ce{Cu-O}$_{\text{Tyr175}}$  & \ce{Cu-O}$_{\text{W229}}$   &  \ce{Cu-O}$_{\text{W230}}$   \\[0.5ex]
\hline \\[-1.5ex]                             
\textbf{1}$^{\text{fixed}}$  &\ce{Cu^{II}}  ($\frac{1}{2}$)    &     2.02           &    2.07                     &  1.98                              &    2.80                       &   2.28                        &  2.11        \\[0.5ex]
\textbf{1}$^{\text{free}}$   &\ce{Cu^{II}}  ($\frac{1}{2}$)    &     2.03           &    2.03                     &  1.99                              &    2.34                       &   2.83                        &  2.03        \\[0.5ex]
\textbf{1}$^{\text{free},a}$ &\ce{Cu^{II}}  ($\frac{1}{2}$)    &     2.02           &    2.02                     &  1.97                              &    2.48                       &   3.00                        &  2.06        \\[0.5ex] 
\textbf{1}$^{\text{free},b}$ & \ce{Cu^{II}} ($\frac{1}{2}$)    &     2.04           &    1.98                     &  2.02                              &    2.47                       &   2.96                        &  2.07        \\[0.5ex]
\textbf{1}\cite{kim2014}     &\ce{Cu^{II}}  ($\frac{1}{2}$)    &     1.99           &    2.08                     &  1.99                              &    3.08                       &      -                        &  2.33        \\[0.5ex]                                 
\textbf{1}\cite{gudmundsson2014}$^{,c}$ & \ce{Cu^{I}} ($\frac{1}{2}$)&     1.99            &   2.07                     & 1.98                               &     -                         & 2.22                          &  2.12       \\[0.5ex] 
\hline \\[-1.5ex]
\textbf{2}$^{\text{fixed}}$    & \ce{Cu^{I}} ($0$)            &     1.97           &    2.11                     &  1.99                              &   2.87                        &   2.30                        & 3.01         \\[0.5ex]
\textbf{2}$^{\text{free}}$     & \ce{Cu^{I}} ($0$)            &     1.95           &    2.08                     &  1.96                              &   3.03                        &   2.28                        & 3.02         \\[0.5ex]
\textbf{2}$^{\text{free},a}$   & \ce{Cu^{I}} ($0$)            &     1.93           &    2.09                     &  1.93                              &   3.05                        &   2.54                        & 3.03         \\[0.5ex]
\textbf{2}$^{\text{free},b}$   &  \ce{Cu^{I}} ($0$)           &     1.97           &    1.97                     &  1.95                              &   2.90                        &   2.74                        & 3.04         \\[0.5ex]
\textbf{2}\cite{kim2014}       &      \ce{Cu^{I}} ($0$)       &     1.93           &    2.14                     &  1.93                              &   4.37                        &   2.19                        &  -           \\[0.5ex]
\textbf{2}\cite{kjaergaard2014}&      \ce{Cu^{I}} ($0$)       &     1.91           &    2.27                     &  1.91                              &   3.23                        &   3.32                        & 3.11         \\[0.5ex]
\textbf{2}\cite{gudmundsson2014}$^{,c}$ & \ce{Cu^{I}} ($0$)      &     1.98           &    2.18                     & 1.98                               &    -                          &    -                          &  -           \\[0.5ex] 
 \hline \\[-1.5ex]  
2YET\cite{quinlan2011}         & \ce{Cu^{I/II}} ($0,\frac{1}{2}$)  &  2.32              &    2.10                     &  2.43                              &   2.80                        &    2.65                       & 2.23         \\[0.5ex]
3ZUD\cite{quinlan2011}         & \ce{Cu^{I/II}} ($0,\frac{1}{2}$)  &  2.03              &    2.20                     &  1.91                              &  2.92                         &   2.89                        &  -           \\[0.5ex]
4EIR\cite{li2012}              & \ce{Cu^{I/II}} ($0,\frac{1}{2}$)  &  1.99              & 2.25                        & 1.92                               &  2.76                         &   -                           & 1.84         \\[0.5ex]
5ACF\cite{frandsen2016}        & \ce{Cu^{I/II}} ($0,\frac{1}{2}$)  &  2.06              & 1.88                        & 2.16                               &  2.47                         &   -                           & -            \\[0.5ex]
4ALC\cite{gudmundsson2014}$^{,c}$ & \ce{Cu^{II}} ($\frac{1}{2}$)      &  1.97              & 2.12                        & 1.99                               & -                             & 2.21                         & 2.19         \\[0.5ex]
4ALT\cite{gudmundsson2014}$^{,c}$ & \ce{Cu^{I}} ($0$)                 &  1.91              & 2.19                        & 1.94                               & -                             &  -                            & -            \\[0.5ex]
\hline \hline 
\end{tabular}
$^a$ Optimized with TPSS-D3/def2-TZVPD $^b$ Optimized with B3LYP-D3/def2-TZVPD.$^c$ This is from an AA10 LPMO enzyme whose reduction in the X-ray beam has been carefully followed.
\end{table*}

We start by discussing the results obtained with TPSS/def2-SV(P) and system 2 relaxed (entries \textbf{1}$^{\text{free}}$ and \textbf{2}$^{\text{free}}$ 
in \ref{opt_rest_state}). 
Despite the reduction of the Cu ion, the three \ce{Cu-N} bonds lengths do not change by more than 0.08 {\AA}, which emphasizes the fact that the active 
site is constructed to accommodate both Cu$^{\text{II}}$ and Cu$^{\text{I}}$.   Further,  
 the effect of reducing Cu$^{\text{II}}$ to Cu$^{\text{I}}$ is an elongation of the Cu--O bonds of Tyr175 and the equatorial water molecule (W230) to 3.0 {\AA}. 
 Meanwhile, the Cu--O bond to the other water molecule shortens from 2.8 to 2.3 {\AA}. Thereby, the Cu ion 
becomes essentially four-coordinated, rather than 5--6-coordinated (octahedral).
 
Considering these small differences in the \ce{Cu-N} bonds lengths upon reduction, it is surprising that the calculated distances reproduce the distances in the starting crystal structure (2YET\cite{quinlan2011}) so poorly: 
Those to the side-chain imidazole rings are 0.3--0.5 {\AA} too short. 
We therefore, optimized structures with the larger def2-TZVPD basis set, both with TPSS and B3LYP functionals. 
However, as can be seen in \ref{opt_rest_state}, this did not lead to any major changes in the Cu--N bond lengths. In particular, those to the His side chains changed by less than 0.05 {\AA}.
Comparing with a range of other crystal LPMO structures 
(also in \ref{opt_rest_state}) shows that the Cu--N$^\delta$ bonds in the 2YET are highly untypical, being 0.3--0.5 {\AA} too long.
If we instead compare with the other crystal structures, the calculated Cu--N bond lengths fall well within the range observed in the crystal structures, with a maximum difference 
of 0.10 {\AA} for the Cu--N$_\text{His1}$ in the oxidized state and 0.05 {\AA} for  the other distances, compared to the average of the five crystal structures.

Comparing structures with the surrounding protein (system 2) fixed or free to relax (fixed or free structures in \ref{opt_rest_state}) shows that there are 
only small differences for the reduced state \textbf{2} (up to 0.03 {\AA}, except for the weak bond to Tyr175, which changes by 0.16 {\AA}). 
However, for the oxidized state (\textbf{1}), much larger changes are seen:
The bond length to Tyr175 decreases from 2.80 to 2.34 {\AA}, whereas that to the axial water molecule increases by almost the same amount, i.e. from 2.28 to 2.83 {\AA}.
This indicates that the crystal structure represents a predominantly reduced state (with a long bond to Tyr175) and that fixing the junction at the C$^\beta$ atoms of Tyr175 is too 
restrictive to model the full flexibility of this residue during the reduction. As will be discussed in Section \ref{CuO2-states}, it is also too restrictive 
to describe the distance in intermediate \textbf{3}. Therefore, we generally  expect large differences for this bond distance when compared to various vacuum 
studies where it is customary to fix C$^{\beta}$ to its initial location from the crystal structure, and this is indeed the case (as will be discussed below).  
The Tyr175 residue has recently been speculated to have implications for the LMPO mechanism\cite{frandsen2016} and the Cu--O distances are 
therefore crucial, but it should also be remembered that axial bonds in Cu complexes are weak and extremely flexible 
(i.e. the distance to Cu can vary much at a minimal expense in energy).\cite{ryde1996b,olsson1999,ryde2000}.
This is reflected in the rather large variation of these bond lengths in the structures optimized with different basis sets and DFT functionals (up to 0.5 {\AA}) and in the various crystal structures (up to 0.7 {\AA}) as can be seen in \ref{opt_rest_state}.

Our structures for \textbf{1} and \textbf{2} are mostly in agreement with the results in previous computational studies\cite{gudmundsson2014,kim2014,kjaergaard2014} (also included in Table 1).
The Cu--N distances agree, except that the Cu--N$_{\text{His1}}$ distance in the reduced state (\textbf{2}) is somewhat shorter (1.97--2.11 \AA\ in our structures, 
compared to 2.14--2.27 \AA\ in the previous studies). 
All studies also suggest that the coordination number of the Cu ion decreases when it is reduced, as expected.
However, the various studies differ in their predictions regarding which of the three prospective O ligands bind and at what distance.
For the oxidized state (\textbf{1}), all studies agree that the equatorial water molecule binds strongly, although the Cu--O distance is $\sim$0.2 \AA\ longer in the study of 
Kim et al.\cite{kim2014} than in the other studies.
Both Kim et al.\cite{kim2014} and Gudmundsson et al.\cite{gudmundsson2014} also suggested that the second water molecule coordinates to Cu (although Ref.~\citenum{kim2014} does not report 
the bond distance to the axial water molecule) in agreement with our result with a fixed surrounding. 
However, when we allow the protein to relax, we instead find that Tyr175 coordinates weakly to Cu, whereas the axial water molecule practically dissociates.

For the reduced state (\textbf{2}), all studies agree that both Tyr175 and the equatorial water molecule effectively dissociate. 
However, for the axial water molecule, the results differ.
Kim et al.\cite{kim2014} suggested that it binds rather strongly at a Cu--O distance of 2.19 \AA, 
whereas the other two studies indicated that it also dissociates.\cite{gudmundsson2014,kjaergaard2014}
Our results are intermediate: With the def2-SV(P) basis set, we obtain a rather short Cu--O bond of $\sim$2.2 \AA. 
However, with the larger def2-TZVPD basis set, and especially with the B3LYP functional, the Cu--O bond becomes appreciably longer, 2.54 and 2.74 \AA, respectively.

Clearly, this reflects the flexibility of weak Cu--O bonds, as mentioned above.\cite{ryde1996b,olsson1999,ryde2000} 
These bond lengths are determined more by interactions with the surrounding protein than by the Cu--O interaction. 
Therefore, it is likely that our QM/MM results, with an explicit account of the surroundings, give the more accurate results. 
However, it is clear that these bonds are sensitive to the theoretical treatment, as our results indicate and as has been pointed out before.\cite{kjaergaard2014}

\subsection{The copper--superoxide intermediates}\label{CuO2-states}

\begin{figure}[!htb]
    \centering
        \includegraphics[scale=0.75]{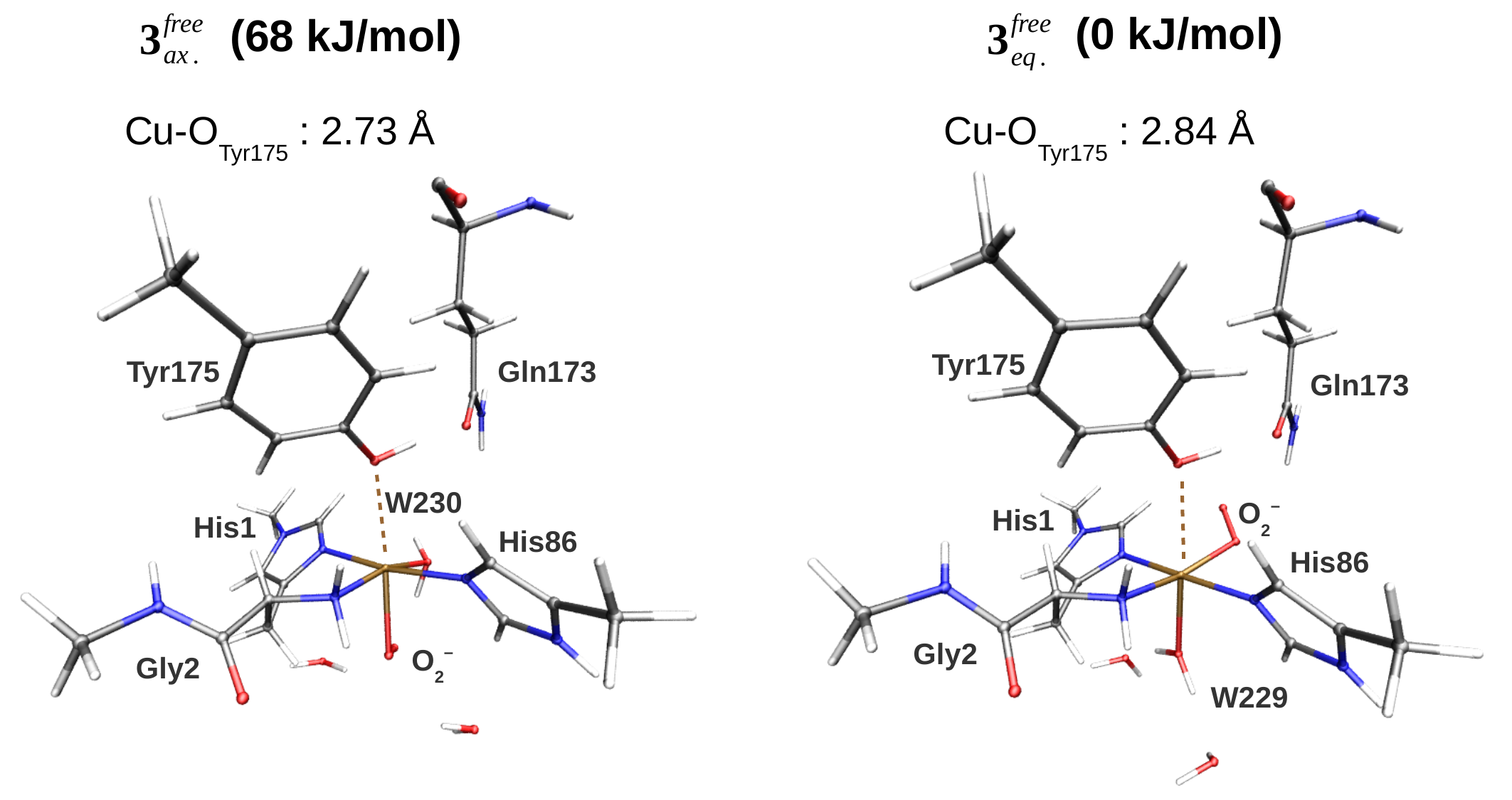}
 \caption{Comparison of the axial and equatorial isomers of \textbf{3}, showing also the Gln173 residue that can interact with \ce{O2-}. The 
structures were optimised in the triplet states ($S=1$) with the TPSS-D3 functional and the def2-SV(P) basis set.
The corresponding bond lengths are shown in \ref{opt_CuOO_state}. \label{opt_CuOO_state_fig}} 
\end{figure}

Next, we discuss the nature of the Cu$^\mathrm{II}$--superoxide adduct, \textbf{3}.
It is expected to form through the binding of \ce{O2} to the reduced active site (cf. \ref{lpmo_active_site}). 
As mentioned above, there are two possible isomers (\textbf{3}$_{\text{eq}}$ or \textbf{3}$_{\text{ax}}$) of this complex, depending on whether \ce{O2} replaces the equatorial or the axial water molecule. 
In variance to the previous studies, we have studied both isomers.
The optimized structures and selected bond-distances of the two isomers are shown in \ref{opt_CuOO_state_fig} and \ref{opt_CuOO_state}, 
whereas the energy difference between them is shown in the upper part of \ref{Cu_o2_ax_vs_eq}.
\begin{table*}[htb!]
\centering
\caption{Cu--ligand bond lengths (\AA) for the active site of LPMO. The results were obtained with the TPSS-D3 functional and the def2-SV(P) basis set, unless otherwise specified. 
The \ce{Cu-O}$_{\text{W}}$ distance corresponds to \ce{Cu-O}$_{\text{W229}}$ for \textbf{3}$_{\text{eq}}$ and \ce{Cu-O}$_{\text{W230}}$ for \textbf{3}$_{\text{ax}}$. \label{opt_CuOO_state} }
\begin{tabular}{lccccccc}
 \hline\hline \\[-2.0ex]
\textbf{Method}                          &  Spin  &   \ce{Cu-N$^{\epsilon}$_{His86}} &  \ce{Cu-N}$_{\text{His1}}$  & \ce{Cu-N}$^{\delta}_{\text{His1}}$ &  \ce{Cu-O}$_{\text{Tyr175}}$  & \ce{Cu-O}$_{2}$   &  \ce{Cu-O}$_{\text{W}}$    \\[0.5ex]%
\hline \\[-1.5ex]  
\textbf{3}$^{\text{fixed}}_{\text{eq}}$  & $S=1$  &    2.06          &     2.15                    &    2.01                              &    2.89                       &  2.04             &    2.24   \\[0.5ex]
\textbf{3}$^{\text{free}}_{\text{eq}}$   & $S=1$  &    2.06          &     2.13                    &    2.00                              &    2.84                       &  2.04             &    2.29   \\[0.5ex]    
\textbf{3}$^{\text{free},a}_{\text{eq}}$ & $S=1$  &    2.06          &     2.12                    &  2.00                                &  2.94                         &  2.01             &    2.40    \\[0.5ex]
\textbf{3}$^{\text{free},b}_{\text{eq}}$ & $S=1$  &    2.08          &     2.11                    & 2.01                                 & 2.84                          & 1.99              &    2.46    \\[0.5ex]                   
Ref.~\citenum{kjaergaard2014}            & $S=1$  &    1.98          &   2.09                      & 1.97                                 &  3.35                         &  1.98             &    3.76    \\[0.5ex]                                      
\hline
\\[-2.0ex]
\textbf{3}$^{\text{fixed}}_{\text{eq}}$  & $S=0$  &   2.06           &     2.15                    &    2.01                              &    2.87                       &  2.02             &    2.24   \\[0.5ex]
\textbf{3}$^{\text{free}}_{\text{eq}}$   & $S=0$  &   2.06           &     2.12                    &    2.00                              &    2.82                       &  2.03             &    2.30    \\[0.5ex]                  
\hline
\\[-2.0ex]
\textbf{3}$^{\text{fixed}}_{\text{ax}}$  & $S=1$  &    2.31          &          2.05               & 2.15                                 &  2.58                         &     2.09          &  2.10   \\[0.5ex]
\textbf{3}$^{\text{free}}_{\text{ax}}$   & $S=1$  &    2.11          &          2.02               & 2.05                                 &  2.73                         &     2.29          &  2.03   \\[0.5ex] 
Ref.~\citenum{kim2014}                   & $S=1$  &    1.98          &          2.18               & 1.98                                 &  3.82                         &     1.96          &  2.33   \\[0.5ex]                                  
\hline \\[-2.0ex]
\textbf{3}$^{\text{fixed}}_{\text{ax}}$  & $S=0$  &    2.31          &          2.04               & 2.14                                 &  2.57                         &     2.10          &  2.08  \\[0.5ex]
\textbf{3}$^{\text{free}}_{\text{ax}}$   & $S=0$  &    2.08          &          2.03               & 2.03                                 &  2.71                         &     2.30          &  2.02  \\[0.5ex] 
Ref.~\citenum{kim2014}                   & $S=0$  &    1.98          &          2.16               & 1.98                                 &   3.80                        &     1.96          &  2.30  \\[0.5ex] 
\hline \hline  \end{tabular}

$^a$ Optimized with TPSS-D3/def2-TZVPD $^b$ Optimized with B3LYP-D3/def2-TZVPD.
\end{table*}
\begin{table*}[htb!]
\centering
\caption{Energy difference $\Delta E = E(\textbf{3}_{\text{ax}})- E(\textbf{3}_{\text{eq}})$ and spin-state splittings $\Delta E =E^{S}({\textbf{3}}) - E^{T}(\textbf{3})$ 
in kJ/mol. $\Delta E_{\text{QM/MM}}$ and $\Delta E_{\text{QM+ptch}}$ are defined in Eq.~\ref{QQ_MM_opt} and $\Delta E_{\text{QM/MM}} = E_{\text{MM123}} - E_{\text{MM1}}$ from the same equation. $E_{\text{QM}}$ 
is the energy of the QM system, without any point-charge model. $E_{\text{big-QM}}$ is the big-QM energy. 
All energies were calculated or extrapolated ($\Delta E_{\text{big-QM}}$) with the def2-TZVPP basis set on structures optimized with 
QM/MM using the def2-SV(P) basis set. In $\Delta E_{\text{big-QM}}$, 10.8 kJ/mol (\textbf{3}$^{\text{free}}$) or $-$1.3 kJ/mol ($\textbf{3}^{\text{free}}_{\text{eq}}$) of the total energy 
is a correction obtained as the difference 
between def2-TZVPP and def2-SV(P) calculations on the QM systems in \ref{qm_systems}. \label{Cu_o2_ax_vs_eq}}
\begin{tabular}{llcccccc}
 \hline\hline \\[-2.0ex]
\textbf{State}                           & Func      & Spin      &   $\Delta E_{\text{QM/MM}}$ &  $ \Delta E_{\text{QM+ptch}}$  &  $\Delta E_{\text{QM}}$ &  $\Delta E_{\text{MM}}$ & $\Delta E_{\text{big-QM}}$ \\[0.5ex]
\hline \\[-1.5ex]   
\textbf{3}$^{\text{fix}}$                & TPSS-D3   & $S=1$     & 63.4                        & 76.5                           &  74.9                   &  $-$4.4                 & -    \\[0.5ex] 
\textbf{3}$^{\text{free}}$               & TPSS-D3   & $S=1$     & 67.8                        & 51.0                           &  65.5                   &  $-$14.6                & 87.7 \\[0.5ex]
\hline    \\[-2.0ex]
$\textbf{3}^{\text{free}}_{\text{ax}}$   & TPSS-D3  & $S=1,0$    & 4.2                         & 5.4                            &  2.7                    & $-$1.2                  & - \\[0.5ex]
$\textbf{3}^{\text{free}}_{\text{ax}}$   & B3LYP-D3  & $S=1,0$   & $-$0.7                        & 0.8                            &  2.8                    & $-$1.2                 & - \\[0.5ex]
$\textbf{3}^{\text{free}}_{\text{eq}}$   & TPSS-D3  &  $S=1,0$   & 12.5                        & 10.6                           &  14.0                   &  2.0	            & 10.5\\[0.5ex]
$\textbf{3}^{\text{free}}_{\text{eq}}$   & B3LYP-D3 &  $S=1,0$   & 13.2                        & 11.2                           &  16.5                   &  2.0                    & - \\[0.5ex] 
\hline \hline  	
\end{tabular}  
\end{table*}

Our results clearly show that the equatorial isomer is most stable. The QM/MM calculations predict an energy difference of 68 kJ/mol.
This is in agreement with the observation in the previous section that the equatorial water molecule effectively dissociates when the oxidized enzyme is reduced, 
providing a free coordinate site to the \ce{O2} molecule.
The point charges, representing the environment, give a contribution of 15 kJ/mol 
(the difference between $\Delta E_{\text{QM+ptch}}$ and $\Delta E_{\text{QM}}$). The electrostatic 
contribution from the protein is thus non-negligible, yet not very large. The contributions 
from the MM force field is of a similar magnitude, but of an opposite sign, and accordingly the vacuum result ($\Delta E_{\text{QM}}$) is close to the 
QM/MM result ($\Delta E_{\text{QM/MM}}$). We additionally carried out a big-QM calculation that included all residues within 5 {\AA} of the active site. 
This gave an energy difference of 77 kJ/mol with the def2-SV(P) basis set. An estimate of the effect of increasing the basis set size to 
def2-TZVPP can be obtained from the difference between def2-SV(P) and def2-TZVPP for the smaller QM system. This effect is 11 kJ/mol, resulting 
in the total difference of 88 kJ/mol given in \ref{Cu_o2_ax_vs_eq}.    

It should be noted that the protein significantly influences the structures.
As for the resting states \textbf{1} and \textbf{2}, 
previous computational studies on \textbf{3} imposed restrictions during the structure optimization (by freezing selected atoms) 
to mimic this protein effect. We compare both isomers with previous calculated results in \ref{opt_CuOO_state}  
and in both cases we find that the structures differ significantly. In particular, we find large differences in the \ce{Cu-O} distance to Tyr175, although it is long in all structures 
(2.57--2.94 \AA\ in our structures, but 3.35--3.82 \AA\ in the previous studies, see Refs.~\citenum{kjaergaard2014,kim2014}).

Both Cu$^{\text{II}}$ and \ce{O2-} have one unpaired electron. These two electrons can either couple ferromagnetically in a triplet state or antiferromagnetically in a singlet state. 
We have studied both states and it can be seen from \ref{opt_CuOO_state}, that the structures for the two spin states are almost identical.
The energy differences between the two spin states are reported in the lower part of \ref{Cu_o2_ax_vs_eq}. 
For \textbf{3}$_{\text{eq.}}$, the triplet is most stable. The energy difference is 13 kJ/mol, both with the TPSS and B3LYP functionals, and the big-QM result is only 2 kJ/mol lower.  
This is in reasonable agreement with 19 kJ/mol
obtained in a previous study\cite{kjaergaard2014}, which is remarkable considering the large differences in the obtained structures. 

 For the axial isomer, the singlet--triplet energy splitting calculated with the QM region and a point-charge model of the environment ($\Delta E_{\text{QM+ptch}}$) are 
5 kJ/mol for TPSS and 1 kJ/mol for B3LYP, which means that the two spin states are essentially degenerate. In fact, 
adding the MM energy of $-1$ kJ/mol (giving $ \Delta E_{\text{QM/MM}}$) is enough to make the singlet the ground state at the B3LYP level. Thus it is not possible 
to settle which of the two spin states is most stable for the axial isomer within the accuracy of current  methods. 
Our spin-splitting energy estimates for $\textbf{3}_{\text{ax}}$ are somewhat lower than the 19 kJ/mol that was obtained in the study by Kim et al.\cite{kim2014}. 

\section{Discussion}\label{discussion}

Both spectroscopic, crystallographic and computational studies have suggested that the 
reduction of \textbf{1} to \textbf{2} is accompanied by a decrease in the coordination number of the Cu atom. This is confirmed in  our calculations, but the 
individual structures for \textbf{1} and \textbf{2} are rather different from those obtained in previous computational studies that have neglected the protein environment. 
We can confirm a previous suggestion that large basis set are required before a significant bond elongation of the axial water molecule (W229) is obtained. However, 
even with large basis sets, we find that this water molecule is still weakly coordinated.   
We also find large differences in our structures of \textbf{3} compared to previous calculations on smaller active site models. Here we highlight that 
\textbf{3}$_{\text{eq}}$ is not four-coordinate, as has been suggested\cite{beeson2015}; instead, our structures indicate that the 
axial water molecule remains coordinated to the Cu ion, although with a rather long distance (2.24--2.46 \AA), the length of which is sensitive to the 
DFT method and the basis set (cf. \ref{opt_CuOO_state}).

\begin{figure}  
\includegraphics[width=0.75\textwidth]{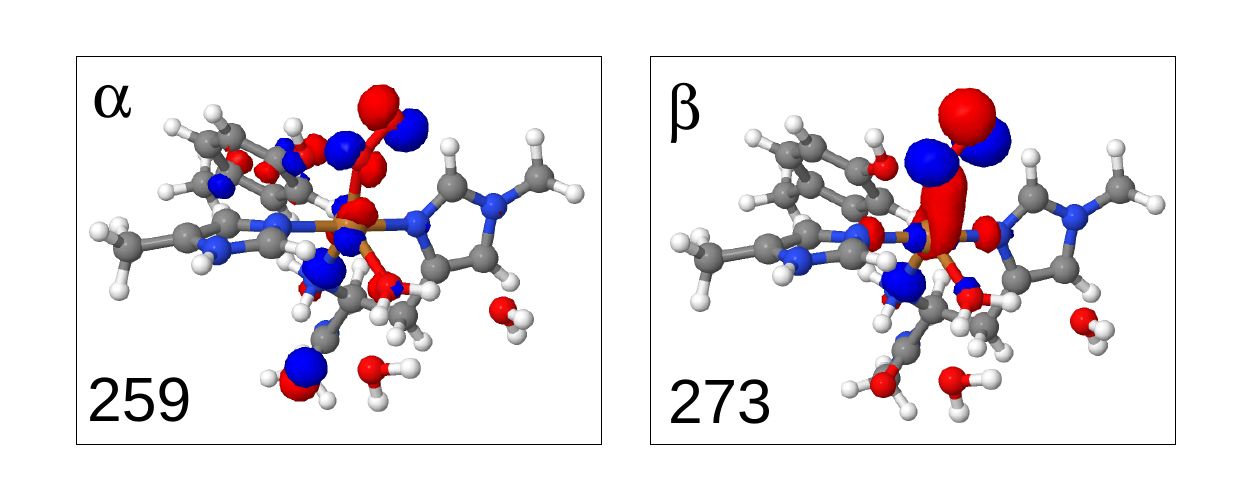}  
\caption{Pair of molecular orbitals for \textbf{3}$_{\text{eq}}$, involved in the Cu $3d_{\pi}$--O$_{\pi}$ interaction with the \ce{O2-} $\pi$ orbitals. 
\label{mo_plots}}   
\end{figure}

In most suggestions for the reaction mechanism of LPMOs, the substrate has been hydroxylated at either C1 or C4\cite{kim2014,beeson2015,phillips2011,beeson2012}, starting 
with a hydrogen abstraction from the substrate.  
The Cu--superoxide complex, \ce{[Cu^{II}O2]+}, is involved either by directly abstracting a hydrogen atom from the substrate\cite{beeson2012,phillips2011} or as precursor 
for an Cu--oxyl radical, which then abstracts the hydrogen\cite{kim2014}. The fact that the superoxide can have both axial and equatorial 
isomers has not been considered in any quantitative studies, although it was noted 
by Beeson et al.,\cite{beeson2015} who suggested that the axial isomer would be unstable, based on a Jahn--Teller distortion argument. 
Our calculations show that the axial isomer \textit{is} stable, but they also confirm and quantify that the the axial isomer is significantly  less 
stable than the equatorial one (the difference is more than 60 kJ/mol). There are two sources of stabilization of the equatorial isomer. 
One is that the equatorial isomer is stabilized by interactions between the superoxide 
and the second-coordination-sphere Gln173 residue (2.24 \AA~away), showing a possible role for this highly conserved residue (cf.~\ref{opt_CuOO_state_fig}). Another source is that 
the equatorial coordination of \ce{O2-} provides better possibility for $\pi$-interaction with the \ce{Cu} 3d-orbitals. The involved orbitals for \textbf{3}$_{\text{eq}}$ 
are shown in \ref{mo_plots}, and more extensive molecular orbital plots for \textbf{3}$_{\text{eq.}}$ and \textbf{3}$_{\text{ax.}}$ are shown 
in Figures S1 and S2 in the supporting information. From these figures it can be seen that 
 stabilizing d$_{\pi}$-O$_{\pi}$ interactions are completely absent for \textbf{3}$_{\text{ax}}$. 

Mechanistic insight has so far been hampered by the lack of crystal structures including a bound substrate, which is difficult to obtain owing 
to the low solubility of cellulose and chitin. 
However, very recently LPMOs that target smaller (soluble) oligosacharides as substrates have been discovered and a crystal structure for a (C4 specific)
LPMO complexed with cellotriose and cellohexose have been reported.\cite{frandsen2016} 
In the LPMO--oligosacharide complex, a \ce{Cl-} ion occupies the equatorial binding site and it is suggested that this is the binding site of 
\ce{O2}, in agreement with our results. Note however, that the 
LPMO reported in Ref.~\citenum{frandsen2016} has a rather different protein scaffold compared to the protein investigated here, and the inclusion of substrate can therefore 
not be carried out by simple means (e.g.~ overlaying C$_{\alpha}$-atoms). In this study, we have rather employed a protein that  is either 
identical or bears close resemblence to those used in previous computational studies, and we can thus more directly compare the computational approaches.    
The large differences in the obtained structures compared to previous computations strongly indicate that an inclusion of the protein matrix is pivotal.   
Although the binding of the substrate may
affect the active site, the large energy difference between axial and equatorial isomers found here indicates that the equatorial superoxide adduct is the active species 
or a precursor. Finally, we should emphasize that we cannot at present state rule out that the active species is a Cu--oxyl complex as suggested in Ref.~\citenum{kim2014}.    

\section{Conclusion}

We have presented the first QM/MM calculations on a LPMO enzyme, based on the AA9 enzyme from \textit{Thermoascus aurantiacus}. 
We investigated the resting state (\textbf{1}), its reduced form (\textbf{2}) and the Cu--superoxide complex (\textbf{3}). 
For all intermediates, the calculated structures are significantly different from those obtained in previous computational studies where the protein environment was neglected. 
For \textbf{3}, there exist two possible isomers, and in this study we have found that one of these (with an equatorial \ce{O2-} ligand) is much more stable than the other (more than 60 kJ/mol). 
Our further studies on the LPMO mechanism will therefore focus 
on this intermediate, or intermediates derived from this. Moreover, 
our future studies will also include the LPMO--substrate complexes, employing the same QM/MM computational protocol as employed here. 
This will allow us  to investigate the reactivity of both the superoxide intermediate (\textbf{3}) and Cu--oxyl complexes.

\section{Computational Methods}

\subsection{Protein setup}\label{setup}

The starting coordinates where taken from the 1.5 {\AA} resolution X-ray 
structure from \textit{Thermoascus aurantiacus}, which belongs to the fungal LPMO (type 3) family.\cite{quinlan2011,beeson2015}. This structure shows the protein in the resting state, possibly 
with a partly reduced active-site Cu ion. The structure is deposited in the protein data bank (2YET) and was also employed in a previous computational study\cite{kim2014}. 
The structure is a dimer that contains 462 amino acids and 625 crystal water molecules, amounting to 4199 atoms in total. 
Here, we consider only the monomer (chain A) and the paper refers to chain A, unless explicitly specified.

The crystal structure contains eight amino acids with alternative conformations, namely 
Asp10, Met25, Ser26, Asn27, Leu41, Ser117, Gln167 and Lys214. We selected the conformation 
with highest occupation or the first conformation if they had the same occupation numbers. 
Hydrogen atoms were added using the Maestro 
protein preparation tools.\cite{Maestro2015} For the titratable residues (2 arginine, 5 lysine, 7 histidine, 14 aspartate and 5 glutamate residues)  
the Maestro program employs the PROPKA program\cite{olsson2011} to estimate pK$_{\text{a}}$ values. The individual residues were 
visually inspected and their solvent exposure and hydrogen-bond network were assessed.
In this study, all Arg and Lys were protonated (+1) 
whereas the Asp and Glu residues were in their carboxylate forms ($-1$). 

The His residues have two possible protonation sites and 
in the following, we denote histidines as HIE (N$^{\epsilon 2}$ protonated), HID (N$^{\delta 1}$ protonated) or HIP (both nitrogens protonated). 
The first (N-terminal) histidine is a special case because the imidazole ring is 
methylated on the N$^{\epsilon 2}$ atom, whereas  N$^{\delta 1}$ coordinates to the Cu ion. For the remaining histidine residues, we employed the protonation states 
HIE57, HID86, HIP87, HID158, HIP164 and HIP201. The N$^{\delta 1}$ atom of HIE57 receives a hydrogen bond from the backbone NH group of Arg58, 
HID86 coordinates to Cu through N$^{\epsilon 2}$, HIP87 forms a salt-bridge from H$^{\delta 1}$ to the carboxylate group of Asp132 and a hydrogen bond 
to a crystal water through H$^{\epsilon 2}$. HID158 forms a hydrogen bond to the carbonyl oxygen of Asn92 through H$^{\delta 1}$, 
whereas the N$^{\epsilon 2}$ atom accepts a hydrogen bond from the side-chain 
indole NH group of Trp79.  HIP164 and HIP201 are solvent exposed on the surface of the protein. 
With this charge assignment, the total charge of the protein in the resting state (\textbf{1}) was $-7$. 

The protein contains four cysteine residues that are cross-linked by disulfide bridges in the pairs Cys56--Cys178 and Cys97--Cys101. 
The carboxy-terminal residue Gly228 was missing from the X-ray analysis and it was left out from the calculations.
The X-ray structure contained one glycerol and four acetate molecules, which were all removed. 

\subsection{RESP charges}  

\begin{figure}[tbh!]
\centering
  \includegraphics[scale=0.60]{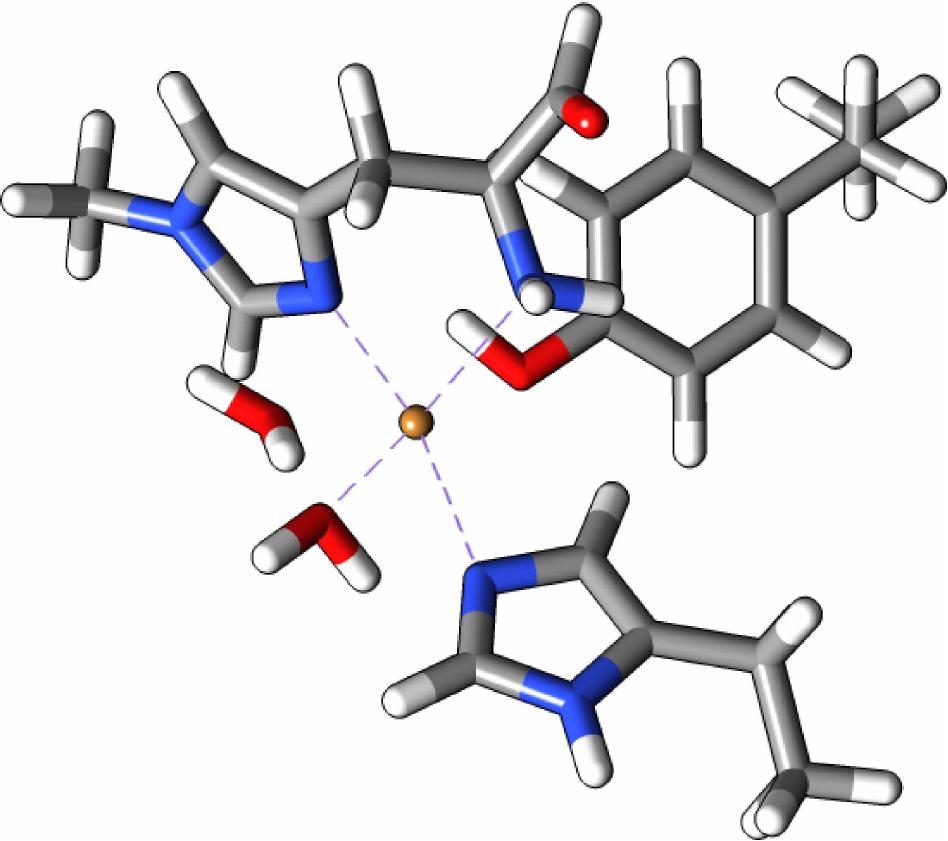}
 \caption{Structure used to obtained RESP charges. }  
\label{resp_rest_stateHopt_for_equi_cut}
\end{figure}

Restrained electrostatic potential (RESP) charges for the metal center and its first coordination sphere were obtained by fitting to the electrostatic potential (ESP). 
The employed structure was taken from the protein (see \ref{resp_rest_stateHopt_for_equi_cut}; it includes all ligands coordinating to the Cu ion in any of the studied complexes) 
and only hydrogen atoms were optimized, employing the TPSS functional\cite{tao2003} together with 
a def2-SV(P) basis set\cite{bs920815sha,eichkorn1997}.  All calculations in this section 
were carried out with a development version of Turbomole 7.0\cite{dm891013abhhk} (modified to write out the ESP points). 
The ESP points were sampled with the Merz--Kollman scheme\cite{singh1984,besler1990} using default radii for all atoms\cite{besler1990} and 2.0 
{\AA} for Cu\cite{sigfridsson1998}. 
They were employed by resp program (a part of the AMBER\cite{amber14} package) to calculate the RESP charges. 

\subsection{Equilibration}  \label{equil}

The system described in Section \ref{setup} was equilibrated by simulated annealing with the AMBER\cite{amber14} software.  
The protein was immersed in a sphere of TIP3P water molecules with a radius of 40 {\AA}, generated by the tLeAP program in the AMBER suite. 
Heavy atoms in the protein and crystal water molecules were kept fixed at their crystal-structure positions. 
During the first 200 ps, the system was heated up to 370 K. This was followed by cooling from 370 K to 0 K over 400 ps. 
The temperature was regulated with the Berendsen thermostat\cite{ryckaert1977} using a time constant that varied during the simulation: 0.2 ps during the first 200 ps, 1.0 ps during the next 200 ps, 0.5 ps during the following 100 ps and 0.05 ps during the last 100 ps, 
leading to a total simulation time of 600 ps. 
The simulations used a time step of 0.5 fs.
Finally, the system was subjected to a 10000-step minimisation.

\subsection{QM/MM calculations}  

The equilibrated system (Section \ref{equil}) was  employed in QM/MM calculations, employing  
the {\sc ComQum} program.\cite{ryde2001,ryde2002} This program combines the QM  software
Turbomole 7.0\cite{dm891013abhhk} and the MM program AMBER 14\cite{amber14}.   
In {\sc ComQum}, the simulated system is divided into  
three subsystems, labeled systems 1, 2 and 3. System 1 is described with a QM method (here DFT). Systems 2 and 3 are both described with an MM force field. 
System 2 is defined as all atoms within 6 {\AA} of any atom in system 1. 
In the following, we use the label ``free'' for calculations in which the coordinates of atoms in system 2 are optimized. In calculations 
labeled ``fixed''  these are kept fixed. System 3 contains the remaining protein and solvent atoms and are always kept fixed at the equilibrated structure.
When there is a bond between systems 1 and 2 (a junction), the hydrogen link-atom approach was employed: The QM region was capped with hydrogen atoms (hydrogen link atoms), 
the positions of which are linearly related to the corresponding carbon atoms (carbon link atoms) in the full system.\cite{ryde1996a,reuter2000} 
 
The total energy of the system is calculated as 
\begin{equation}
 E_{\text{QM/MM}} = E_{\text{QM+ptch}} + E_{\text{MM123}} - E_{\text{MM1}} . \label{QQ_MM_opt}
\end{equation}
$E_{\text{QM+ptch}}$ is the QM energy of system 1, including hydrogen link atoms and a point-charge model of systems 2 and 3 
(with point charges taken from the Amber force field and excluding only the carbon link atoms).\cite{hu2011} $E_{\text{MM123}}$ is the total MM energy of the full 
system (but with the charges of the QM system zeroed) and $E_{\text{MM1}}$ is the MM energy of system 1 (still with zeroed charges). 
\begin{figure}[tbh!]
\centering
  \includegraphics[scale=0.75]{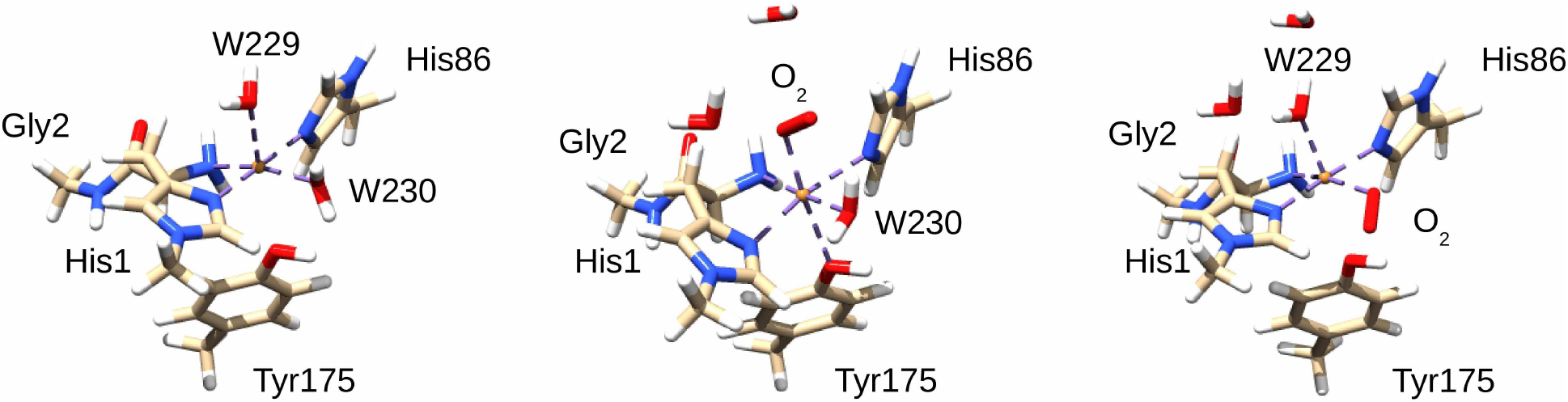}
 \caption{The QM systems employed for \textbf{1}, \textbf{3}$_{\text{ax}}$ and \textbf{3}$_{\text{eq}}$. Model \textbf{2} employed the same QM system as 
\textbf{1} (but with a reduced copper ion) and the same QM system is employed 
for triplet and singlet variants of \textbf{3}. \label{qm_systems} }  
\end{figure}

In our study, we focused on the three states \textbf{1}, \textbf{2} and \textbf{3} in \ref{lpmo_active_site}. For \textbf{3} we considered 
two possible isomers with the superoxide ion binding in either the axial or the equatorial position, denoted \textbf{3}$_{\text{ax}}$ and  \textbf{3}$_{\text{eq}}$.  
Both isomers of \textbf{3} may attain either a triplet state or an antiferromagnetically coupled (open-shell) singlet state. The latter was calculated as a broken-symmetry state\cite{noodlemann1986}. 

The QM region always included the first coordination sphere i.e.~the imidazole ring of His86 and the phenol ring of Tyr175, both 
capped with a hydrogen atom (replacing C$^{\alpha}$). The whole terminal histidine residue was included, as well as parts of the next amino-acid (Gly2; junction at the carbonyl C atom). 
The QM regions are shown in \ref{qm_systems}. 
  
For both isomers of \textbf{3}, we included two solvent molecules in the QM region, in addition to the water molecule coordinating 
to Cu$^{\text{II}}$. This decision was made based on initial calculations, in which they were absent. The initial calculations showed that the energy difference 
between the isomers were dominated by 
van der Waals energy from the MM calculations (amounting to over 80 kJ/mol). 
By decomposing the energy into contributions from individual residues, it was shown that the large change in van der Waals energy almost exclusively was caused  
by two solvent water molecules. To shift this large energy contribution to the more accurate QM part of the energy, we decided to include 
these two solvent molecules in the QM system. As expected, this lowered the energy change associated with the MM part to around 17 kJ/mol.    

In our initial setup, we noted that
the side chain of Gln173 was flipped by the Maestro preparation tools. It is located in the second coordination sphere at hydrogen-bonding distance of both Tyr175 and one of the Cu-bound water molecules (2.7--2.8 \AA\ distance between the heavy atoms). 
Therefore, we decided to perform an optimization both in the flipped (\textbf{1a}) 
and original (\textbf{1b}) conformation. The latter conformation was employed in the study by Kim et al.\cite{kim2014} which was performed in vacuum, but included 
parts of the second coordination sphere. 
In conformation \textbf{1b} and the crystal structure, the side-chain carbonyl group of Gln173 can receive a hydrogen bond from the Tyr175 OH group and another from the Cu-bound water molecule. 
In the other conformation, the side-chain \ce{NH2} group instead donates a hydrogen bond to Tyr175, but it cannot donate any hydrogen bond to the water molecule.
Therefore, \textbf{1b} is favored by about 30 kJ/mol, even when calculated on a structure equilibrated for the \textbf{1a} state where the 
hydrogen bonding networks around the Gln173 \ce{NH2} group are somewhat unfavorable for \textbf{1b}.  
We will therefore focus on the conformation in \textbf{1b} from this point on. To ensure that the hydrogen network around the Gln173  \ce{NH2} was sufficiently 
relaxed, the structure was re-equilibrated, and the QM/MM optimization of \textbf{1} was 
carried out from this re-equilibrated structure.  Starting structures for \textbf{2} and \textbf{3} were built 
from \textbf{1}, by reducing the charge of the QM region (\textbf{2}) and replacing W229 (\textbf{3}$_{\text{ax}}$) or W230 (\textbf{3}$_{\text{eq}}$) 
with \ce{O2}. 

The structure optimizations employed the def2-SV(P) basis set\cite{bs920815sha,eichkorn1997}, and the dispersion-corrected TPSS-D3 functional\cite{tao2003,grimme2010} 
with Becke--Johnson damping\cite{grimme2011}. All reported energies are obtained from these structures by single-point calculations with the more accurate 
def2-TZVPP basis set\cite{bs920815sha} and including the entire protein (Systems 2 and 3), represented by point charges 
(the effect of increasing the basis set is usually around 1 kJ/mol for energy differences).  
In the case of \textbf{3}, we also tried the B3LYP/def2-TZVPP combination (with the same 
structures).  
It has previously been emphasised that basis sets of at least triple-zeta quality are required to model the structure of \textbf{2}\cite{kjaergaard2014}.  
For states \textbf{1}, \textbf{2} and \textbf{3}$_{\text{eq}}$ we therefore probed the quality of the TPSS-D3/def2-SV(P) structures by increasing 
the basis set to def2-TZVPD or replacing the TPSS-D3 functional with B3LYP-D3\cite{becke1988,becke1993,lee1988}. For both \textbf{1} and \textbf{2}, the basis-set effect is 
significant for the \ce{Cu-O} bonds (as we will discuss in more detail in Section \ref{results}). 
For \textbf{3}, the effect is smaller, and here def2-SV(P) basis set is sufficient to obtain reliable QM/MM structures.  

It should be emphasized that relaxation of system 2 was found to have a significant influence on the obtained structure, also within the first coordination sphere. 
Consequently, we have 
in general focused on the results obtained with a relaxed system 2. The effect of relaxing system 2 is shortly discuss for intermediates 
\textbf{1} and \textbf{2} in Section \ref{rest_state}. Otherwise, results with a fixed system 2 are included in the Tables for comparison, but not thoroughly discussed. 
  
\subsection{Big-QM calculations}

The use of a point-charge model for the environment can be somewhat inaccurate and for other metalloenzymes it has been advocated to improve the QM/MM energies with 
single-point energy calculations with larger QM systems based on the QM/MM optimized structures.\cite{hu2013,sumner2013}. Following 
Refs.~\citenum{hu2013,sumner2013} we have therefore defined a large QM system comprised of all residues within 5.0 {\AA} of the active site shown in \ref{lpmo_active_site}. 
In addition, junctions were moved at least three residues away from the active site and we included the only two buried charged residues in the protein, Arg157 and Glu159, 
which form a salt bridge, rather close to the active site (Glu159 was actually included already with the 5 {\AA} criterium). 
The total system had a charge of +3 and was comprised of 628 atoms. Around this system we employed a conductor-like screening model (COSMO)\cite{klamt1993} 
with a dielectric constant of $\epsilon=4.0$.
The calculations employed the TPSS method and the def2-SV(P) basis set, based on structures obtained with the same specifications 
(note that this is sufficient for \textbf{3}, as described in previous section). 
The big-QM energy was enhanced with a DFT-D3 dispersion correction, 
calculated for the same big-QM region with Becke--Johnson damping, third-order terms and default parameters for the TPSS functional. 
Finally, the energies were extrapolated to the def2-TZVPP basis set using two QM calculations on the normal QM system.   

\begin{acknowledgement}


This investigation has been supported by grants from the Swedish research council (project 2014-5540) and from COST through Action CM1305 (ECOSTBio). 
The computations were performed on computer resources provided by the Swedish National Infrastructure for Computing (SNIC) at Lunarc (Lund University). 
E.D.H. thanks the Carlsberg Foundation for a postdoctoral fellowship (Grant no. CF15-0208). 
\end{acknowledgement}


\begin{thebibliography}{10}

\bibitem{klemm2005}
Klemm,~D.; Heublein,~B.; Fink,~H.-P.; Bohn,~A. \emph{Angew. Chem. Int. Ed.}
  \textbf{2005}, \emph{44}, 3358--3393.

\bibitem{chang2007}
Chang,~M. \emph{Curr. Opin. Chem. Biol.} \textbf{2007}, \emph{11}, 677--684.

\bibitem{himmel2007}
Himmel,~M.~E.; Ding,~S.-Y.; Johnson,~D.~K.; Adney,~W.~S.; Nimlos,~M.~R.;
  Brady,~J.~W.; Foust,~T.~D. \emph{Science} \textbf{2007}, \emph{315},
  804--807.

\bibitem{hemsworth2013a}
Hemsworth,~G.~R.; Davies,~G.~J.; Walton,~P.~H. \emph{Curr. Opin. Struc. Biol.}
  \textbf{2013}, \emph{23}, 660--668.

\bibitem{beeson2015}
Beeson,~W.~T.; Vu,~V.~V.; Span,~E.~A.; Phillips,~C.~M.; Marletta,~M.~A.
  \emph{Annu. Rev. Biochem.} \textbf{2015}, \emph{84}, 923--946.

\bibitem{span2015}
Span,~E.~A.; Marletta,~M.~A. \emph{Curr. Opin. Struct. Biol.} \textbf{2015},
  \emph{35}, 93--99.

\bibitem{harris2014}
Harris,~P.~V.; Xu,~F.; Kreel,~N.~E.; Kang,~C.; Fukuyama,~S. \emph{Curr. Opin.
  Chem. Biol.} \textbf{2014}, \emph{19}, 162--170.

\bibitem{quinlan2011}
Quinlan,~R.~J. et~al. \emph{Proc. Nat. Sci. USA} \textbf{2011}, \emph{108},
  15079--15084.

\bibitem{raguz1992}
Raguz,~S.; Yag{\"u}e,~E.; Wood,~D.~A.; Thurston,~C.~F. \emph{Gene}
  \textbf{1992}, \emph{119}, 183--190.

\bibitem{armesilla1994}
Armesilla,~A.~L.; Thurston,~C.~F.; Yag{\"u}e,~E. \emph{FEMS Microbiol. Lett.}
  \textbf{1994}, \emph{116}, 293--299.

\bibitem{karkehabadi2008}
Karkehabadi,~S.; Hansson,~H.; Kim,~S.; Piens,~K.; Mitchinson,~C.; Sandgren,~M.
  \emph{J. Mol. Biol.} \textbf{2008}, \emph{383}, 144--154.

\bibitem{harris2010}
Harris,~P.~V.; Welner,~D.; McFarland,~K.~C.; Re,~E.; Poulsen,~J.-C.~N.;
  Brown,~K.; Salbo,~R.; Ding,~H.~Vlasenko,~E.; Merino,~S.; Xu,~F.; Cherry,~J.;
  Larsen,~S.; Leggio,~L.~L. \emph{Biochemistry} \textbf{2010}, \emph{49},
  3305--3316.

\bibitem{dotson2007}
Dotson,~W.~D.; Greenier,~J.; Ding,~H. \emph{{Polypeptide from a cellulolytic
  fungus having cellulolytic enhancing activity}}, US Patent 7361495 B2, 2007.

\bibitem{brown2008}
Brown,~K.; Harris,~P.; Zaretsky,~E.; Edward~Re.,~D.; Vlasenko,~D. Alfredo Lopez
  de~Leon,~D. \emph{Polypeptide from a cellulolytic fungus having cellulolytic
  enhancing activity}, US Patent Appl. 11/046, 124, 2008.

\bibitem{vaaje-kolstad2010}
Vaaje-Kolstad,~G.; Westereng,~B.; Horn,~S.~J.; Liu,~Z.; Zhai,~H.~S{\o}rlie,~M.;
  Eijsink,~V. G.~H. \emph{Science} \textbf{2010}, \emph{330}, 219--222.

\bibitem{levasseur2013}
Levasseur,~A.; Drula,~E.; Lombard,~V.; Coutinho,~P.~M.; Henrissat,~B.
  \emph{Biotechnol Biofuels} \textbf{2013}, \emph{6}, 41.

\bibitem{phillips2011}
Phillips,~C.~M.; Beeson,~W.~T.; Cate,~J.~H.; Marletta,~M.~A. \emph{ACS Chem.
  Biol.} \textbf{2011}, \emph{6}, 1399--1406.

\bibitem{beeson2012}
Beeson,~W.~T.; Phillips,~C.~M.; Cate,~J. H.~D.; Marletta,~M.~A. \emph{J. Am.
  Chem. Soc.} \textbf{2012}, \emph{134}, 890--892.

\bibitem{horn2012}
Horn,~S.~J.; Vaaje-Kolstad,~G.; Westereng,~B.; Eijsink,~V. G.~H.
  \emph{Biotechnol Biofuels} \textbf{2012}, \emph{5}, 45--56.

\bibitem{forsberg2011}
Forsberg,~Z.; Vaaje-Kolstad,~G.; Westereng,~B.; Bun{\ae}s,~A.~C.;
  Stenstr{\o}m,~Y.; MacKenzie,~A.; S{\o}rlie,~M.; Horn,~S.~J.; Eijsink,~V.
  G.~H. \emph{Protein Sci.} \textbf{2011}, \emph{20}, 1479--1483.

\bibitem{hemsworth2013b}
Hemsworth,~G.~R.; Taylor,~E.~J.; Kim,~R.~Q.; Gregory,~R.~C.; Lewis,~S.
  J.~Turkenburg,~J.~P.; Parkin,~A.; Davies,~G.~J.; Walton,~P.~H. \emph{J. Am.
  Chem. Soc.} \textbf{2013}, \emph{135}, 6069--6077.

\bibitem{vaaje-kolstad2012}
Vaaje-Kolstad,~G.; B{\o}hle,~L.~A.; G{\aa}seidnes,~S.; Dalhus,~B.;
  Bj{\o}r{\aa}s,~M.; Mathiesen,~G.; Eijsink,~V. G.~H. \emph{J. Mol. Biol.}
  \textbf{2012}, \emph{416}, 239--254.

\bibitem{aachmann2012}
Aachmann,~F.~L.; S{\o}rlie,~M.; Skj{\aa}k-Br{\ae}k,~G.; Eijsink,~V. G.~H.; ;
  Vaaje-Kolstad,~G. \emph{Proc. Nat. Sci. USA} \textbf{2012}, \emph{109},
  18779--18784.

\bibitem{vaaje-kolstad2013}
Vaaje-Kolstad,~G.; Horn,~S.~J.; S{\o}rlie,~M.; Eijsink,~V. G.~H. \emph{FEBS J.}
  \textbf{2013}, \emph{280}, 3028--3049.

\bibitem{solomon2014}
Solomon,~E.~I.; Heppner,~D.~E.; Johnston,~E.~M.; Ginsbach,~J.~W.; Cirera,~J.;
  Qayyum,~M.; Kieber-Emmons,~M.~T.; Kjaergaard,~C.~H.; Hadt,~R.~G.; Li~Tian,~L.
  \emph{Chem. Rev.} \textbf{2014}, \emph{114}, 3659--3853.

\bibitem{westereng2011}
Westereng,~B.; Ishida,~T.; Vaaje-Kolstad,~G.; Wu,~M.; Eijsink,~V. G.~H.;
  Igarashi,~K.; Samejima,~M.~St{\aa}hlberg,~J.; Horn,~S.~J.; Sandgren,~M.
  \emph{PLoS ONE} \textbf{2011}, \emph{6}, e27807.

\bibitem{li2012}
Li,~X.; Beeson~IV,~W.~T.; Phillips,~C.~M.; Marletta,~M.~A.; Cate,~J. H.~D.
  \emph{Structure} \textbf{2012}, \emph{20}, 1051--1061.

\bibitem{bey2013}
Bey,~M.; Zhou,~S.; Poidevin,~L.; Henrissat,~B.; Coutinho,~P.~M.; Berrin,~J.-G.;
  Sigoillot,~J.-C. \emph{Appl. Environ. Microbiol.} \textbf{2013}, \emph{79},
  488--496.

\bibitem{wu2013}
Wu,~M.; Beckham,~G.~T.; Larsson,~A.~M.; Ishida,~T.; Kim,~S.; Payne,~C.~M.;
  Himmel,~M.~E.; Crowley,~M.~F.; Horn,~S.~J.; Westereng,~B.; Igarashi,~K.;
  Samejima,~M.; St{\aa}hlberg,~J.; Eijsink,~V. G.~H.; Sandgren,~M. \emph{J.
  Biol. Chem.} \textbf{2013}, \emph{288}, 12828--12839.

\bibitem{sommerhalter2005}
Sommerhalter,~M.; Lieberman,~R.~L.; Rosenzweig,~A.~C. \emph{Inorg. Chem.}
  \textbf{2005}, \emph{44}, 770--778.

\bibitem{gudmundsson2014}
Gudmundsson,~M.; Kim,~S.; Wu,~M.; Ishida,~T.; Momeni,~M.~H.; Vaaje-Kolstad,~G.;
  Lundberg,~D.; Royant,~A.; St{\aa}hlberg,~J.; Eijsink,~V. G.~H.;
  Beckham,~G.~T.; Sandgren,~M. \emph{J. Biol. Chem.} \textbf{2014}, \emph{289},
  18782--18792.

\bibitem{tian2009}
Tian,~C.; Beeson,~W.~T.; Iavarone,~A.~T.; Sun,~J.; Marletta,~M.~A.; Cate,~J.
  H.~D.; Glass,~N.~L. \emph{Proc. Natl. Acad. Sci. U.S.A.} \textbf{2011},
  \emph{106}, 22157--22162.

\bibitem{berka2011}
Berka,~R.~M. et~al. \emph{Nat. Biotechnol.} \textbf{2011}, \emph{29}, 922--927.

\bibitem{forsberg2014a}
Forsberg,~Z.; MacKenzie,~A.; S{\o}rlie,~M.; R{\o}hr,~{\AA}.~K.; Helland,~R.;
  Arvai,~A.~S.; Vaaje-Kolstad,~G.; Eijsink,~V. G.~H. \emph{Proc. Sci. Nat. USA}
  \textbf{2014}, \emph{111}, 8446--8451.

\bibitem{forsberg2014b}
Forsberg,~Z.; R{\o}hr,~{\AA}.~K.; Mekasha,~S.; Anderson,~K.~K.; Eijsink,~V.
  G.~H.; Vaaje-Kolstad,~G.; S{\o}rlie,~M. \emph{Biochemistry} \textbf{2014},
  \emph{53}, 1647--1656.

\bibitem{hemsworth2014}
Hemsworth,~G.~R.; Henrissat,~B.; Davies,~G.~J.; Walton,~P.~H. \emph{Nat. Chem.
  Biol.} \textbf{2014}, \emph{10}, 122--126.

\bibitem{vu2014a}
Vu,~V.~V.; Beeson,~W.~T.; Span,~E.~A.; Farquhar,~E.~R.; Marletta,~M.~A.
  \emph{Proc. Natl. Acad. Sci. U.S.A.} \textbf{2014}, \emph{111}, 13822--13827.

\bibitem{leggio2015}
Leggio,~L.~L. et~al. \emph{Nat. Commun.} \textbf{2015}, \emph{6}, 1--9.

\bibitem{vu2014b}
Vu,~V.~V.; Beeson,~W.~T.; Phillips,~C.~M.; Cate,~J. H.~D.; Marletta,~M.~A.
  \emph{J. Am. Chem. Soc.} \textbf{2014}, \emph{136}, 562--565.

\bibitem{borisova2015}
Borisova,~A.~S.; Isaksen,~T.; Dimarogona,~M.; Kognole,~A.~A.; Mathiesen,~G.;
  V\'{a}rnai,~A.; R{\o}hr,~{\AA}.~K.; Payne,~C.~M.; S{\o}rlie,~M.;
  Sandgren,~M.; Eijsink,~V. G.~H. \emph{J. Biol. Chem.} \textbf{2015},
  \emph{290}, 22955--22969.

\bibitem{kjaergaard2014}
Kjaergaard,~C.~H.; Qayyum,~M.~F.; Wong,~S.~D.; Xu,~F.; Hemsworth,~G.~R.;
  Walton,~D.~J.; Solomon,~E.~I. \emph{Proc. Natl. Acad. Sci. USA}
  \textbf{2014}, \emph{111}, 8797--8802.

\bibitem{kim2014}
Kim,~S.; St{\aa}hlberg,~J.; Sandgren,~M.; Patond,~R.~S.; Beckham,~G.~T.
  \emph{Proc. Nat. Sci. USA} \textbf{2014}, \emph{111}, 149--154.

\bibitem{hu2009}
Hu,~L.; Eliasson,~J.; Heimdal,~J.; Ryde,~U. \emph{J. Phys. Chem. A}
  \textbf{2009}, \emph{113}, 11793--800.

\bibitem{hu2011}
Hu,~L.; S{\"o}derhjelm,~P.; Ryde,~U. \emph{J. Chem. Theory Comput.}
  \textbf{2011}, \emph{7}, 761--777.

\bibitem{hu2013}
Hu,~L.; S{\"o}derhjelm,~P.; Ryde,~U. \emph{J. Chem. Theory Comput.}
  \textbf{2013}, \emph{9}, 640--649.

\bibitem{sumner2013}
Sumner,~S.; S\"{o}derhjelm,~P.; Ryde,~U. \emph{J. Chem. Theory Comput.}
  \textbf{2013}, \emph{9}, 4205--4214.

\bibitem{karasulu2013}
Karasulu,~B.; Patil,~M.; Thiel,~W. \emph{J. Am. Chem. Soc.} \textbf{2013},
  \emph{135}, 13400--13413.

\bibitem{quesne2014}
Quesne,~M.~G.; Latifi,~R.; Gonzalez-Ovalle,~L.~E.; Kumar,~D.; de~Visser,~S.~P.
  \emph{Eur. J. Chem.} \textbf{2014}, \emph{20}, 435--446.

\bibitem{finkelmann2014}
Finkelmann,~A.~R.; Senn,~H.~M.; Reiher,~M. \emph{Chem. Sci.} \textbf{2014},
  \emph{5}, 4474--4482.

\bibitem{hedegard2015c}
Hedeg{\aa}rd,~E.~D.; Kongsted,~J.; Ryde,~U. \emph{Angew. Chem. Int. Ed.}
  \textbf{2015}, \emph{54}, 6246--6250.

\bibitem{cortopassi2015}
Cortopassi,~W.~A.; Simion,~R.; Honsby,~C.~E.; Fran\c{c}a,~T. C.~C.;
  Paton,~R.~S. \emph{Eur. J. Chem.} \textbf{2015}, \emph{21}, 18983--18992.

\bibitem{warshel1976}
Warshel,~A.; Levitt,~M. \emph{J. Mol. Biol.} \textbf{1976}, \emph{103},
  227--249.

\bibitem{senn2009}
Senn,~H.~M.; Thiel,~W. \emph{Angew. Chem. Int. Ed.} \textbf{2009}, \emph{48},
  1198--1229.

\bibitem{visser2014}
de~Visser,~S.~P.; Quesne,~M.~G.; Martin,~B.; Comba,~P.; Ryde,~U. \emph{Chem.
  Comm.} \textbf{2014}, \emph{50}, 262--282.

\bibitem{ryde2016}
Ryde,~U. In \emph{{Methods Enzymol.}};
\newblock Elsevier, 2016, Vol. 577, Chapter~6, pp 119--158.

\bibitem{frandsen2016}
Frandsen,~K. E.~H. et~al. \emph{Nat. Chem. Biol.} \textbf{2016}, \emph{12},
  298--305.

\bibitem{ryde1996b}
Ryde,~U.; Olsson,~M. H.~M.; Pierloot,~K.; Roos,~B.~O. \emph{J. Mol. Biol.}
  \textbf{1996}, \emph{261}, 586--596.

\bibitem{olsson1999}
Olsson,~M. H.~M.; Ryde,~U. \emph{J. Biol. Inorg. Chem.} \textbf{1999},
  \emph{4}, 654--663.

\bibitem{ryde2000}
Ryde,~U.; Olsson,~M. H.~M.; Roos,~B.~O.; Pierloot,~K.; De~Kerpel,~J. O.~A.
  \emph{J. Biol. Inorg. Chem.} \textbf{2000}, \emph{5}, 565--574.

\bibitem{Maestro2015}
 \emph{Maestro version 10.2 (2015)}, Schr{\"o}dinger, LLC, New York, NY, 2015.

\bibitem{olsson2011}
Olsson,~M. H.~M.; S{\o}ndergard,~C.~R.; Rostkowski,~M.; Jensen,~J.~H. \emph{J.
  Chem. Theory Comput.} \textbf{2011}, \emph{7}, 525--537.

\bibitem{tao2003}
Tao,~J.; Perdew,~J.~P.; Staroverov,~V.~N.; Scuseria,~G.~E. \emph{Phys. Rev.
  Lett.} \textbf{2003}, \emph{91}, 146401.

\bibitem{bs920815sha}
Sch{\"a}fer,~A.; Horn,~H.; Ahlrichs,~R. \emph{J. Chem. Phys.} \textbf{1992},
  \emph{97}, 2571--2577.

\bibitem{eichkorn1997}
Eichkorn,~K.; Weigend,~F.; Treutler,~O.; Ahlrichs,~R. \emph{Theor. Chem. Acc.}
  \textbf{1997}, \emph{97}, 119--124.

\bibitem{dm891013abhhk}
Ahlrichs,~R.; B{\"a}r,~M.; H{\"a}ser,~M.; Horn,~H.; K{\"o}lmel,~C. \emph{Chem.
  Phys. Lett.} \textbf{1989}, \emph{162}, 165--169.

\bibitem{singh1984}
Singh,~C.~A.; Kollman,~P.~A. \emph{J. Comp. Chem.} \textbf{1984}, \emph{5},
  129--45.

\bibitem{besler1990}
Besler,~B.~H.; Merz~Jr.,~K.~M.; Kollman,~P.~A. \emph{J. Comp. Chem.}
  \textbf{1990}, \emph{11}, 431--439.

\bibitem{sigfridsson1998}
Sigfridsson,~E.; Ryde,~U. \emph{J. Comp. Chem.} \textbf{1998}, \emph{19},
  377--395.

\bibitem{amber14}
Case,~D.~A. et~al. \emph{{AMBER 14}}, 2014, University of California, San
  Francisco.

\bibitem{ryckaert1977}
Ryckaert,~J.-P.~Ciccotti,~G.; Berendsen,~H. J.~C. \emph{J. Comp. Phys.}
  \textbf{1977}, \emph{23}, 321--341.

\bibitem{ryde2001}
Ryde,~U.; Olsen,~M. H.~M. \emph{Int. J. Quantum Chem.} \textbf{2001},
  \emph{81}, 335--347.

\bibitem{ryde2002}
Ryde,~U.; Olsen,~L.; Nilson,~K. \emph{J. Comput. Chem.} \textbf{2002},
  \emph{23}, 1058--1070.

\bibitem{ryde1996a}
Ryde,~U. \emph{J. Comput. Aided Mol. Des.} \textbf{1996}, \emph{10}, 153--164.

\bibitem{reuter2000}
Reuter,~N.; Dejaegere,~A.; Maigret,~B.; Karplus,~M. \emph{J. Phys. Chem. A}
  \textbf{2000}, \emph{104}, 1720--1735.

\bibitem{noodlemann1986}
Noodleman,~L.; Davidson,~E.~R. \emph{Chem. Phys.} \textbf{1986}, \emph{109},
  131--143.

\bibitem{grimme2010}
Grimme,~S.; Antony,~J.; Ehrlich,~S.; Krieg,~H. \emph{J Chem. Phys.}
  \textbf{2010}, \emph{132}, 154104.

\bibitem{grimme2011}
Grimme,~S.; Ehrlich,~S.; Goerigk,~L. \emph{J. Comput. Chem.} \textbf{2011},
  1456--1465.

\bibitem{becke1988}
Becke,~A.~D. \emph{Phys. Rev. A} \textbf{1988}, \emph{38}, 3098--3100.

\bibitem{becke1993}
Becke,~A.~D. \emph{J. Chem. Phys.} \textbf{1993}, \emph{98}, 5648--5652.

\bibitem{lee1988}
Lee,~C.; Yang,~W.; Parr,~R.~G. \emph{Phys. Rev. B} \textbf{1988}, \emph{37},
  785--789.

\bibitem{klamt1993}
Klamt,~A.; Sch{\"u}{\"u}rmann,~G. \emph{J. Chem. Soc., Perkin Trans.}
  \textbf{1993}, \emph{2}, 799--805.

\end{thebibliography}

\newcommand{\Aa}[0]{Aa}
\providecommand{\refin}[1]{\\ \textbf{Referenced in:} #1}

\includepdf[pages={-}]{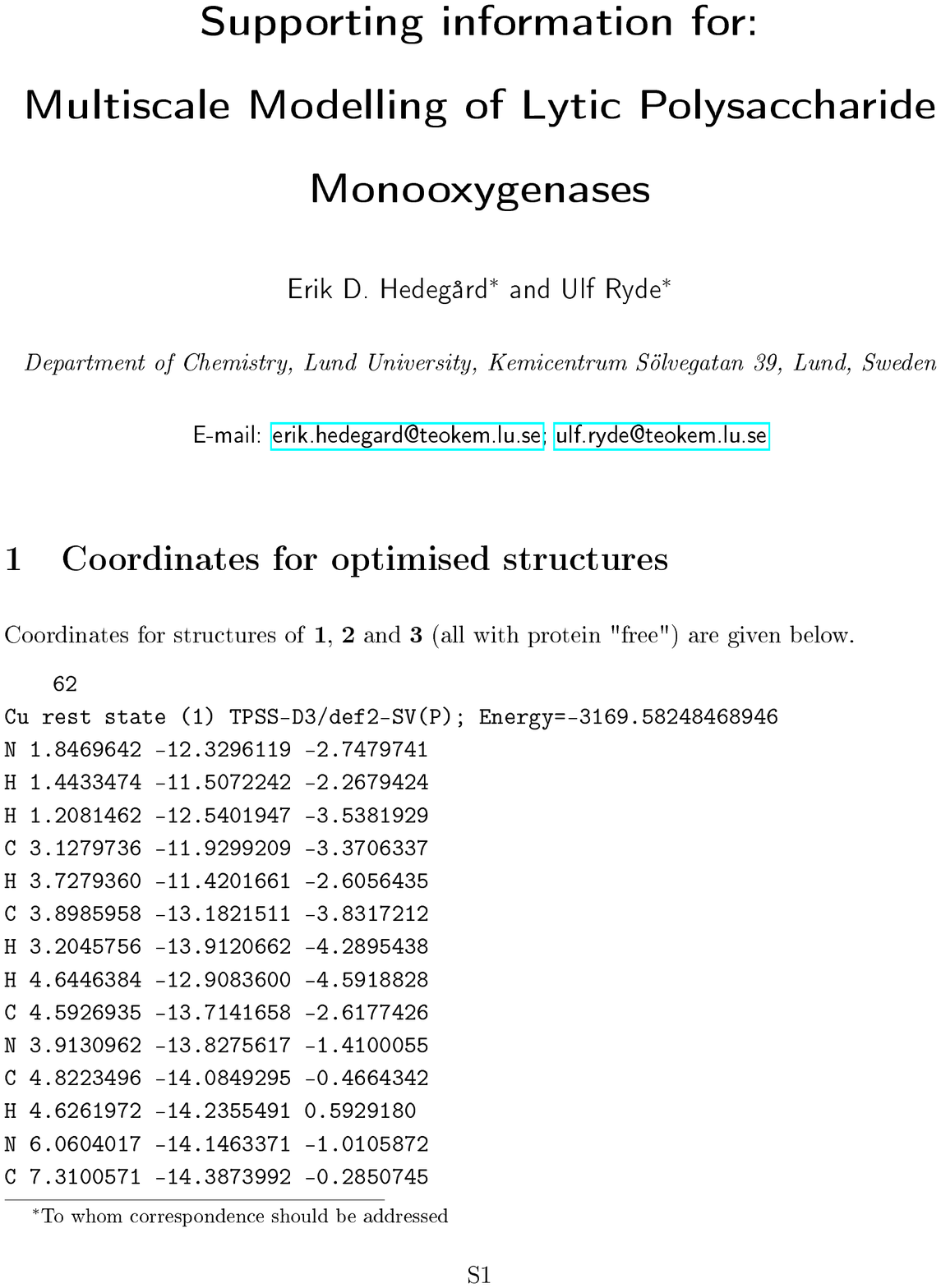}

\end{document}